# Failure of Griffith Theory on Prediction of Theoretical Strength of Ideal Materials

Zhao Liu and Biao Wang*

School of Physics and Sino-French Institute of Nuclear Engineering and Technology, Sun Yat-sen University, Zhuhai 519082, China

Corresponding author: B. Wang, wangbiao@mail.sysu.edu.cn

**Abstract**

Ever since its publication, the Griffith theory is the most widely used criterion for estimating the ideal strength and fracture strength of materials depending on whether the materials contain cracks or not. A Griffith strength limit of ~$E/9$ is the upper bound for ideal strengths of materials. With the improved quality of fabricated samples and the power of computational modeling, people have recently reported the possibility of exceeding the ideal strength predicted by the Griffith theory. In this study, a new strength criterion was established based on the stable analysis of thermodynamical systems; then first-principles density functional theory (DFT) is used to study the ideal strength of four materials (diamond, $c$-BN, Cu, and $CeO_2$) under uniaxial tensile loading along the [100], [110], and [111] low-index crystallographic directions. By comparing the ideal strengths between DFT results and the Griffith theory, it is found that the Griffith theory fails in all the four materials. Further analysis of the fracture mechanism demonstrates that the failure of the Griffith theory is because the ideal strength point does not correspond to creating of new surfaces, which is against the Griffith assumption that the crack intrinsically exists all the time; the failure point corresponds to the crack at the start of propagation.

## 1. Introduction

The theoretical strength of ideal materials is the upper strength limit that a perfect crystal can withstand at 0 K [1, 2]. In reality, it is extremely difficult for practical materials to reach the theoretical ideal strength due to the presence of internal defects and flaws [3-5]. Nowadays, people are trying to fabricate materials that can approach their ideal strengths by reducing size and dimension [6-10]. How to determine the ideal strength of materials is an important issue since it is crucial for understanding the failure of a material under a specific loading mode. In 1921, Griffith [11] considered the failure problem of a material specimen with a crack. He proposed that if the released elastic energy accompanying the unit length increase of the crack was equal to the created surface energy, the specimen with the crack would fail, which made him become a pioneer of fracture mechanics. He also thought that for perfect crystals the space between the atoms played the same role as a tiny crack, and he established a criterion of $\sigma_s = \sqrt{\dfrac{2E\gamma_s}{\pi a}} \sim E/9$ to estimate the ideal strength of materials, where $E$ is the elastic modulus of the materials under uniaxial tensile loading, $a$ is half of the space between the neighboring atoms, and $\gamma_s$ is the specific surface energy. Subsequent developments by Polanyi, Orowan, and other scientists considered the theoretical ideal strength to be $\sim E/10$ [12-14]. Some recent experimental and simulation results for different materials show that there is possibility that the ideal strength of materials may exceed the theoretical strength estimated by the above-mentioned Griffith criterion [15-18]. These findings strongly suggest that the Griffith theory on the prediction of theoretical strength may not be proper for many kinds of materials [19]. In this work, we will show that it is true that the Griffith theory failed to predict theoretical strength since no new surface will be created at the failure point of the perfect materials. On the other hand, to determine the fracture strength of materials when cracks exist, the Griffith equation is the one that has been most widely used and verified by extensive experimental data. In the past two decades, first-principles simulations have proved to be powerful to study the ideal strength of materials because there is no empirical parameters involved [20-24]. However, there is large discrepancy between the DFT (density functional theory) calculated results and experimental results [25]. In fact, in the field of solid mechanics, scientists used to use empirical criteria to determine the material strength. Recently, Wang [26, 27] considered the loaded material as a thermodynamic system, and derived the rational strength criterion by establishing the stable condition of the system. In this study, a new strength criterion was developed to determine the ideal strength of materials. Then, DFT simulations were done for the mechanical behavior of four representative materials, *i.e.*, diamond, copper, *c*-BN, and $CeO_2$, to verify the validity of the established criterion. Further analysis of the DFT results elucidated the reason for the possibility that the experimental ideal stress may exceed the theoretical value given by the Griffith equation, and the inapplicability of the Griffith theory for theoretical strength prediction.

## 2. A thermodynamic approach to determine the theoretical strength of materials

Consider a material specimen as a thermodynamic system under some external mechanical loading in some environmental temperature. To establish a theory to determine the strength of the material, one can imagine that under the quasi-static loading process. For an equilibrium state, if the state is stable under any perturbation, the loaded material specimen has not reached the failure point, whereas, if the state is not stable under some perturbation, the material specimen has reached the critical failure state, and the critical conditions at this state is the strength criterion. This is our basic idea for how to establish the strength criterion to determine the ideal strength for materials. For the first-principle simulation, one can find the maximum stress under uniaxial loading for a given material, but under complex loadings, a strength criterion must be used in any calculations. In what follows, we will show how to establish stable conditions for the material system.

To judge if a state of a thermodynamic system is stable or not, one can add some small perturbations or fluctuations to the system's thermodynamic quantities;

according to the thermodynamic principles, if the system with the perturbation is driven back to its equilibrium state, the state is stable. However, if the system with the perturbation is driven away from its equilibrium state, the state is not stable. Thus, one can establish the stable condition, or the strength criterion.

As stated in modern thermodynamics [28], the entropy reaches its maximum value at equilibrium for an isolated system, which means that any perturbation from the stable equilibrium state can only reduce the entropy, or the state of equilibrium is stable to any perturbation that results in a decrease of the system in entropy. If the system is subjected to different external conditions, the stable condition will change according to thermodynamics. For our case, the material specimen is under the action of applied loading $\sigma$. The first and second law of thermodynamics gives

$$dG = dU - dW = -TdS_i \leq 0 \quad (\sigma, T = \text{constant}), \tag{1}$$

where $G$, $U$, and $W$ are the Gibbs free energy, internal energy, and the work on the system by the applied load, respectively. Equation (1) dictates that at equilibrium, the Gibbs free energy should reach its minimum value by keeping the applied load and temperature constant. The fluctuation in the strain along the boundaries, of which the Gibbs free energy is a function, is quantified by its magnitudes, such as $\delta\varepsilon$, and can be expanded as a power series of the strain perturbation, so we have

$$G = G_{ex} + \delta G + \frac{1}{2}\delta^2 G + \cdots. \tag{2}$$

In the expansion, the term $\delta G$ represents the first-order terms containing $\delta\varepsilon$, $\delta^2 G$ represents the second order terms containing $(\delta\varepsilon)^2$, and so on. At an equilibrium state, the first-order terms are zero, and the leading contribution of the change in Gibbs free energy is due to the second terms. If the second-order variations $\delta^2 G > 0$, the equilibrium state is stable, and if $\delta^2 G \leq 0$, the state is not stable. In such way, one can establish the stable criterion or the strength criterion using the following equation:

$$\delta^2 G = 0, \tag{3}$$

and $\delta G=0$ gives

$$\sigma = \frac{\partial U}{\partial \varepsilon}..\qquad(4)$$

Thus, equation (3) gives the critical value of the strain, and equation (4) gives the critical stress at failure. Generally speaking, the stable condition, equation (3), can be expressed in some explicit form for different problems.

For a penetrating crack with length 2a in a plate with thickness B under the action of tensile loading $\sigma$, the elastic Gibbs free energy can be written in the form as

$$G(a) = -\frac{2\pi\sigma^2 a^2 B}{E'} + 4\gamma(a)Ba, \quad E' = \begin{cases} E/(1-v^2), & plane\ strain \\ E, & plane\ stress \end{cases},\qquad(5)$$

where $\gamma$ is the surface energy density. If one assumes that $a = a_c$ is an equilibrium state, to investigate its stability behavior, the perturbation of the free energy can be expressed in the form

$$G(a_c + \Delta a) = G(a_c) + \left.\frac{\partial G}{\partial a}\right|_{a=a_c}\Delta a + \left.\frac{\partial^2 G}{\partial a^2}\right|_{a=a_c}(\Delta a)^2 + \cdots$$
$$\left.\frac{\partial G}{\partial a}\right|_{a=a_c} = 0, \quad \text{for any equilibrium state}$$
$$\left.\frac{\partial^2 G}{\partial a^2}\right|_{a=a_c} < 0, \text{ for unstable state}\qquad(6)$$

From the first derivative of the free energy being zero, Griffith derived

$$\sigma_c = \sqrt{\frac{2E'\gamma}{\pi a_c}}\ .\qquad(7)$$

In fact, for his case, the surface energy density is a constant, and the second derivative is always negative. Therefore, if the load reaches the value given by equation (7), the system will not be stable, and the specimen will fail. To determine the theoretical strength of an ideal material, Griffith thought that the lattice distance played the same role as the crack length. In this way, equation (7) also gives the theoretical strength. For many ideal materials, the failure point does not correspond to create new surface but corresponds to inducing the lattices rearrangement and electron cloud density to change, as shown below by first-principle calculations. For such cases, the Griffith equation on theoretical strength is not correct. In fact, for those failure mechanisms with different energy dissipation processes in addition to creating a new surface, one needs to set the second derivative at zero to determine the critical crack length, then use the first derivative at zero to determine the critical load.

### 3. The first principle calculation of some materials

To verify the validity of the strength criterion established in the previous section, first-principles simulations were used to study the mechanical behaviors of four different ideal materials under uniaxial tensile loading. The materials chosen in this work are diamond, *c*-BN, Copper, and $CeO_2$. Diamond and c-BN are chosen as examples here is because they are known as the hardest and second-hardest materials existing in nature [29]; Cu is chosen because it is a typical metal [2, 20], while $CeO_2$ is chosen because it is one of the most widely used catalytic metal-oxides [30, 31]. All these materials possess a cubic crystal structure, so uniaxial tensile loading along the three important crystallographic directions ([100], [110], and [111]) are investigated. The DFT calculation methods can be found in the Supplementary Information (SI).

Figure 1 shows the DFT calculated stress-strain relations for the four materials along different uniaxial tensile loading directions. The data for diamond and *c*-BN are in good agreement with previously reported results [32, 33], while the mechanical properties of Cu and $CeO_2$ under different uniaxial loading directions are reported for the first time. The general feature in Figure 1 shows that there is strong anisotropy of the maximum tensile stresses in all the four materials, with the highest ideal stress being along the [100] direction. The results for diamond and *c*-BN are also consistent with previous experimental and simulation work that the cleavage of them would probably occur in the <111> planes due to the markedly lower strength. It is also noted that both materials exhibit isotropic stress response when the strain is lower than 0.1, which indicates that the elastic model in engineering can give a good description for these materials when the strain is small. The isotropic stress response for small strain is not observed in $CeO_2$ possibly because its bonds are ionic. Among the four materials, diamond possesses the highest ideal strength along [100] direction (220 GPa), followed by *c*-BN and CeO$_2$, while Cu possesses the lowest ideal strength of 22 GPa, owing to its metallic characteristics.

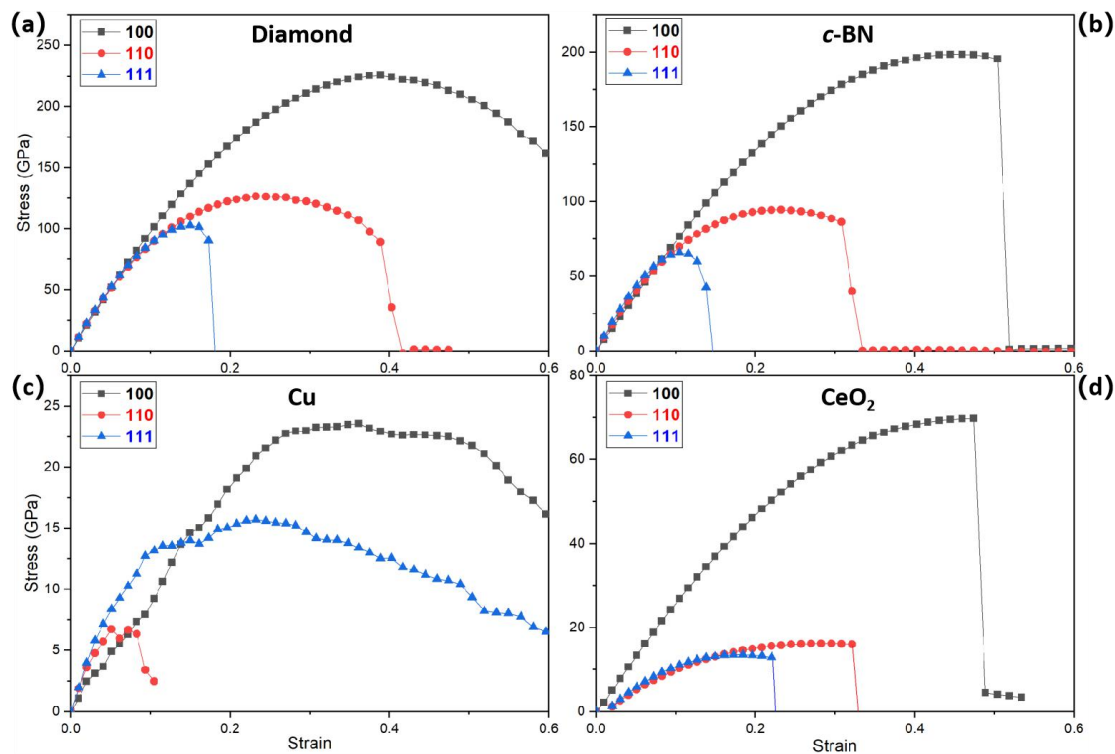

**Figure 1.** The calculated stress-strain relations under different uniaxial tensile loading for (a) diamond; (b) *c*-BN; (c) Cu; and (d) CeO$_2$.

Figure 2 shows the relation between the system energy and stress as a function of the applied tensile strain for the four materials. It can be seen that the energy of the system increases with the increase of applied strain, while the stress keeps increasing until the maximum stress and then decreases. The energy increases because of the structural distortion caused by the tensile strain, whereas the stress decrease after the maximum is attributed to the softening of the bonds in the materials. This can be seen in Figure S5 and S6, where the electron localization function (ELF) of CeO$_2$ shows the evolution of the Ce-O bonds during the uniaxial tensile loading process. By calculating the second derivative of the system energy against the applied strain, the inflection point on the energy curve matches exactly with the ideal stress, which verifies the validity of the criterion as expressed in Eq. 3 and 4. This is confirmed to be true by all the four materials studied, as shown in Figure 2.

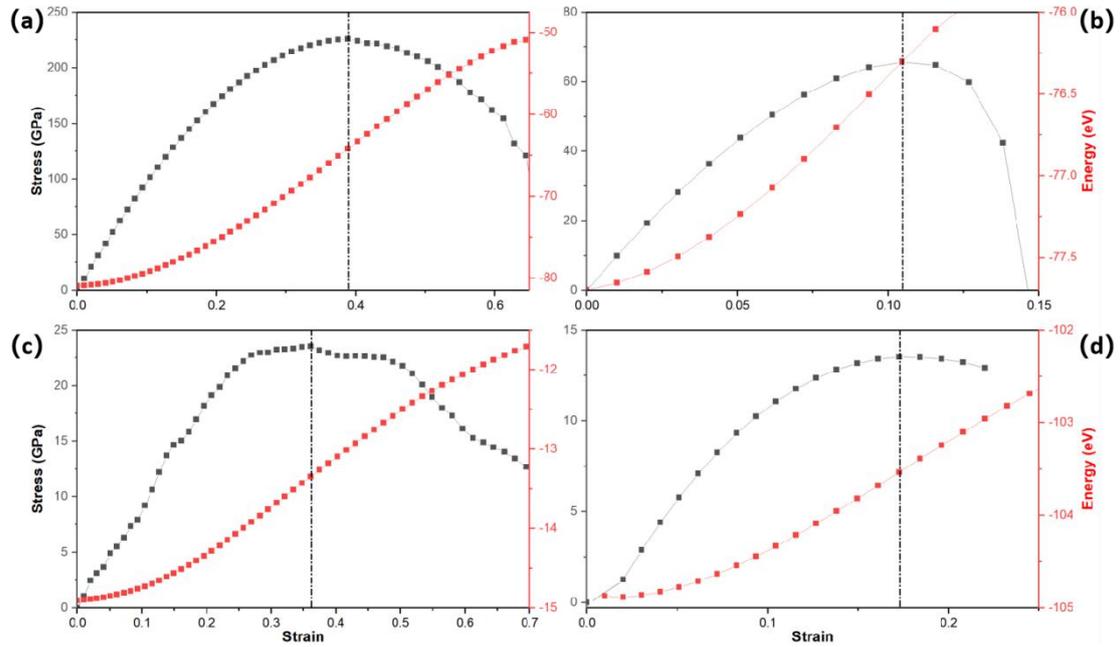

**Figure 2.** The relations between the system energy and stress as functions of the applied strain for (a) diamond along [100] direction; (b) *c*-BN along [111] direction; (c) Cu along [100] direction; and (d) $CeO_2$ along [111] direction.

Recently, with the increasing quality of fabricated ideal materials, people have reported the possibility that the ideal strength of experimental samples may exceed the theoretical values predicted by the Griffith theory. This issue is addressed in the following section. Table 1 shows the discrepancy between the ideal strengths calculated by the Griffith theory and DFT simulations for the four materials. The values predicted by Griffith theory are calculated according to Eq. 7, where the '*a*' is now the lattice parameter of the unit cell instead of the crack length, while the ideal strengths for DFT are determined according to the stress-strain curve in Figure 1. According to Table 1, it can be seen that the DFT ideal strengths along the [100] direction for all the four materials are higher than the values calculated by the Griffith theory, especially for diamond, where the DFT strength is almost 80% higher than that of the Griffith Strength; the discrepancies for c-BN, Cu, and $CeO_2$ are 19.2%, 4.3%, and 30.2%, respectively. Combining our results with the findings of previous works, it is possible that the ideal strength of materials can exceed the value predicted by the

Griffith theory.

**Table 1.** Discrepancy between the ideal strengths calculated by the Griffith theory and determined by DFT results for the four materials.

| Material | Tensile Direction | DFT Strength (GPa) | Griffith Strength (GPa) | Discrepancy |
|---|---|---|---|---|
| Diamond | 100 | 223 | 124 | **+79.8%** |
|  | 110 | 126 | 119 | **+5.9%** |
|  | 111 | 102 | 106 | -3.7% |
| $c$-BN | 100 | 199 | 167 | **+19.2%** |
|  | 110 | 95 | 120 | -20.8% |
|  | 111 | 66 | 159 | -58.8% |
| Cu | 100 | 24 | 23 | **+4.3%** |
|  | 110 | 7 | 29 | -75.8% |
|  | 111 | 16 | 30 | -46.7% |
| $CeO_2$ | 100 | 69 | 53 | **+30.2%** |
|  | 110 | 16 | 35 | -54.3% |
|  | 111 | 13 | 27 | -51.9% |

To explain the possible reason for the failure of Griffith theory, the uniaxial tensile loading along the [110] direction on diamond and $CeO_2$ are taken as examples. Figure 3a shows the average bond length of C-C with the increasing tensile strain along the [110] direction as well as the corresponding stress evolution, and the fractured structure is also shown when the applied strain is 0.403. It can be seen that the fracture occurs after the maximum stress. The bond length of C-C increases gradually before the fracture. When the strain reaches 0.388, the bond length increases sharply from 2.12 to 3.26 Å, indicating the occurrence of crack and fracture. The abrupt increase of the C-C bond length also indicates the softening of the bond. In the case of $CeO_2$ as shown in Figure 3b, the Ce-O bond length increases linearly with the uniaxial strain. When the stress abruptly drops, no significant increase in the bond length is observed, which indicates the rigidity of the Ce-O bond. In Griffith theory, the existence of cracks is considered when estimating the ideal strength. However, it is seen from Figure 3 that the crack does not occur before the ideal stress (peak stress). This may explain the reason for the failure of Griffith theory in some cases. Due to the anisotropy of crystals, it is highly possible that the realistic ideal strength of materials in experiments can exceed the theoretical values calculated by the Griffith theory. Therefore, new strength theory is needed to estimate the ideal strength of materials.

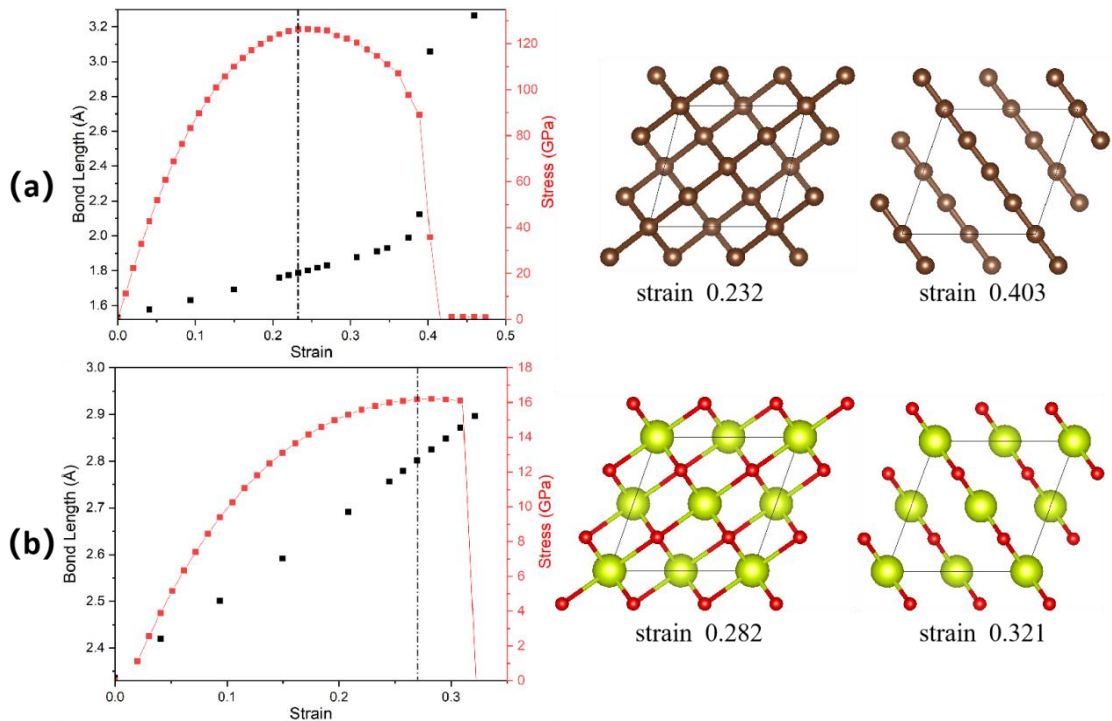

**Figure 3.** (a) The average bond length of C-C with increasing tensile strain along the [110] direction and the corresponding stress curve. The unit cell and atomic positions of strain at 0.232 (ideal stress) and 0.403 (fracture) are shown. (b) The average bond length of Ce-O with increasing tensile strain along the [110] direction, and the corresponding stress curve. The unit cell and atomic positions of strain at 0.282 (ideal stress) and 0.321 (fracture) are also shown.

## 4. Conclusions

In summary, we have established a criterion to determine the ideal strength of materials by the approach of thermodynamics. The criterion suggests that the inflection point of the second derivative of the system energy as a function of strain matches exactly with the ideal stress. First-principles simulations were conducted to study the mechanical behaviors of three example materials, which verify the validity of the criterion. More importantly, we found that there was big discrepancy between the ideal strength calculated by DFT and the Griffith theory, despite the anisotropy of crystals. Further analysis elucidated the reason for the discrepancy is that the material does not contain any cracks when reaching its maximum stress, whereas the Griffith theory considers the existence of cracks even before the fracture occurs. This is supported by some experimental observations that the ideal strength of materials can exceed the theoretical strength estimated by the Griffith theory. Our study demonstrated that the Griffith theory used to predict ideal strength of materials could fail in many cases, and a new form of theory is needed to predict the ideal strength of materials.

## Declaration of competing interest

The authors declare no competing financial interests.

## Acknowledgements

This work was supported by the National Natural Science Foundation of China (Grant Nos. 11832019, 11472313, 13572355, 12002402) and the Fundamental Research Funds for the Central Universities (20lgpy186).